# Electron Transport in Very Clean, As-Grown Suspended Carbon Nanotubes

Jien Cao, Qian Wang and Hongjie Dai*

*Department of Chemistry and Laboratory for Advanced Materials, Stanford University, Stanford, CA 94305, USA*

**Single walled carbon nanotubes (SWNT) have displayed a wealth of quantum transport phenomena thus far.[1-11] Defect free, unperturbed SWNTs with well-behaved or tunable metal contacts are important to probing the intrinsic electrical properties of nanotubes. Meeting these conditions experimentally is nontrivial due to numerous disorder and randomizing factors. Here we show that ~ 1 μm long fully suspended SWNTs grown-in-place between metal contacts afford SWNT devices exhibiting well-defined characteristics over much wider energy ranges than nanotubes pinned on substrates. Various low temperature transport regimes in true-metallic, small and large bandgap semiconducting nanotubes are observed including quantum states shell-filling, -splitting and -crossing in magnetic fields for medium conductance devices. The clean transport data reveals a correlation between the contact junction resistance and the various transport regimes in SWNT devices. Further, we show that electrical transport data can be used to probe the band structures of nanotubes including nonlinear band dispersion.**

Our suspended nanotube devices consisted of individual SWNTs grown[12] in place across pre-formed W/Pt electrodes over a trench with local metal gate at the bottom of the trench (Fig.1a). Device fabrication was as described previously.[9,13] The nanotubes were suspended between Pt contacts (suspended length in the range of $L$~0.5 to 2 μm, Fig. 1b) in native as-grown forms without receiving any wet chemical treatments.

Due to their relative abundance, we first focused on small-bandgap ($E_g$ <100meV) semiconducting (or quasi-metallic) SWNTs.[14] Device #1 (Dev1) comprised of an $L$~500nm suspended SWNT exhibits conductance $G≤1.5\ e^2/h$ in the p-channel and slightly lower $G≤1\ e^2/h$ in the n-channel at $T$=0.3 K (Fig. 1c). The p-channel ($V_g$<-2V) exhibits slow Fabry Perot[5]-like humps with 4 peaks overriding on each hump. The gap region shows $G$~0 and the n-channel exhibits Coloumb blockade (CB)-like peaks (Fig. 1c). Deep into the n-channel (for $V_g$>~ 4V), 4-peak structures (i.e., 3 short period in $G$-$V_g$ peaks followed by 1 longer period, as marked in Fig. 1c top axis) are discerned again but not over any obvious FP-hump. The 4-peak structures correspond to 'shell-filling' of longitudinally quantized states in the SWNT.[7,10,11] For each shell with a given wave-vector $k_n$=$n\pi/L$ ($n$=integer, index for shell), 4 states are available for electron filling due to degeneracy of K and K' subbands and spin degeneracy (Fig. 1d).

A new observation by the current work is a striking evolution pattern of the shell-filling structures in an axial magnetic field **B** (Fig. 2). In the p-channel of Dev1, we observed that under increasing |**B**|, each shell (circled region in Fig. 2a at B=0) splits into two groups with each group of two peaks shifting up or down respectively. At higher B fields, the two up-moving peaks re-group with two down-moving groups from the neighboring shell, forming a new 4-peak shell structure at B≠0 (Fig. 2a circled region away from B=0). The shell-splitting, -crossing and re-forming give rise to well-behaved zigzag patterns in the $G$-$V_g$-$B$ plot. The shell-crossing field is $V_g$ dependent (as marked by the dashed line in Fig. 2a) and scales roughly linearly with $V_g$. That is, for energies closer to the band edge, the shells cross at a lower |**B**| fields.

In Figure 3, we show another suspended quasi-metallic SWNT (Dev2, tube length $L$~700nm) with p-channel conductance $G \leq 1\ e^2/h$. The overall characteristics of this suspended SWNT are similar to that of Dev1 with shell-filling structures over a wide range of energy (or $V_g$) and clear shell-crossing zigzag patterns in **B** fields. The shell-crossing is more rapid than Dev1 with multiple crossings observed in the **B**=0 to 8 T field range (or steeper slope of the dashed line in Fig. 3b than in Fig. 2a). As discussed later, the origin of this difference is the larger bandgap of the SWNT in Dev2 than in Dev1.

The electron transport data above suggests that the suspended SWNTs with $L$< 1μm appear 'ideal' with clean transport signatures over wide energy ranges and magnetic fields. No signs of significant defects, disorder or perturbation to the SWNTs are apparent. We observed that ~70-80% of $L$< 1μm suspended SWNT devices (out of 40-50 quasi-metallic ones) behave this way with well-defined $I$-$V_g$ characteristics over a wide range of $V_g$. In contrast, we have measured numerous SWNTs grown by the same chemical vapor deposition method on substrates but have never obtained as clean transport data as the suspended tubes shown here. It is possible that for SWNTs lying on a substrate, the substrate-SWNT interaction could mechanically flatten the tube,[15] which could introduce slight perturbation to the nanotube electronic property.

For the 4-electron shell structures, we observe similar peak spacings for the 4-peaks within each shell followed by a larger spacing to the next shell (see Fig. 1c and Fig. 2a bar heights along top-axis). We model the three similar peaks by a parameter $U_{eff}$ that is an effective 1-electron addition or charging energy (with exchange, subband mismatch energy in a previous model[7,16] included) for 4-electrons within a shell and the larger spacing for adding another electron to the next shell (Fig.1e) by $U_{eff}+\Delta_n$ where $\Delta_n=E(k_{n+1})-E(k_n)$, $k_n=n\pi/L$ (n=integer) and $E(k)$ is the band dispersion curve (Fig.2c). Using this simple model, we extract $U_{eff}$ from shell-filling 4-peak patterns in $G$-$V_g$. We find that the same nanotube in Dev1 exhibits $U_{eff}$=2.8meV in the p-channel with medium-conductance of $G\sim 1.5\ e^2/h$ (Fig. 1c negative $V_g$ side) and a higher addition energy of $U_{eff}$=3.9 meV in the lower conductance n-channel (Fig.1c positive $V_g$ side).

The well-defined shell–crossing patterns for Dev1&2 (Fig.2a&3b) reflect the electronic band structures $E(k)$ of the two nanotube and $E(k)$ evolution in magnetic field. Specifically, in zero field, the quantization energy of each shell $k_n$ relates to the $E(k)$ dispersion through $E(k_n)=\eta v_F\sqrt{(\Delta k_\perp)^2+k_n^2}$ in which $\Delta k_\perp$ provides a measure of the bandgap ($E_g=\eta v_F\Delta k_\perp$, $v_F$ ~ Fermi velocity in graphene). Due to non-linear $E(k)$, the energy spacing $\Delta_n$ between the quantized shells is energy or $V_g$ dependent and is small near the bandgap due to flatness of the band there (Fig.1d). The energy of each state in a certain shell can then be written as $E_{n,l}=E(k_n)+l\cdot U_{eff}$, in which $n$ is the index of the shells, and $l = 0,1,2,3$ is the index of the four electrons in each shell. We have

$$E_{n,l}=E(k_n)+l\cdot U_{eff}=\eta v_F\sqrt{(\Delta k_\perp)^2+\left(\frac{\pi}{L}\cdot n\right)^2}+l\cdot U_{eff} \qquad (1)$$

The effect of the **B** field can be included by adding the Aharanov-Bohm term[8,9,17,18] $\Delta k_{AB}=\frac{\pi\cdot r}{h/e}B$ (r is tube radius),

$$E_{n,l,B}=E_B(k_n)+l\cdot U_{eff}=\eta v_F\sqrt{\left(\Delta k_\perp\pm\frac{\pi\cdot r}{h/e}B\right)^2+\left(\frac{\pi}{L}\cdot n\right)^2}+l\cdot U_{eff} \qquad (2)$$

where the sign "+" is for K sub-band and "-" for K' sub-band. This gives rise to splitting of K and K' subbands (Fig. 2c) and two up- and down-moving states (Fig.2a circled states) respectively in each shell.

Without the 1-electron addition energy $U_{eff}$ in Eq. 2, calculated energies of the states in each shell and their evolution in B field are shown in Fig.2d (dashed lines for K and K' states) with bandgap $E_g = \eta v_F \Delta k_\perp$ as a fitting parameter. Since each electron takes the lowest energy state, crossing of the states occurs to result in the zigzag pattern of energy levels (solid lines in Fig. 2d) along the **B** field direction. This is the origin of shell-crossing pattern observed experimentally. With the charging energy $U_{eff}$ and 2-fold spin degeneracy included, we derive the electronic states for all 4-states in each shell under various **B** fields). By matching the calculated states (Fig. 2b) with that of experiment (Fig. 2a) including details such as the slopes of the shell-crossing fields, we extract the bandgap $E_g$ of the SWNT. For the SWNTs in Dev1 (Fig. 2) and Dev2 (Fig.3), we obtain $E_g \sim$ 60meV and $\sim$ 100meV respectively.

The physics underlying the more rapid shell-crossing along the field axis for a larger bandgap SWNT (in Dev2 than in Dev1) is the flatter *E(k)* dispersion for the larger bandgap tube near the band edge. This corresponds to a slower group velocity along the tube axis. The slower moving electrons accumulate more AB fluxes while orbiting along the tube circumference,[8,9] therefore experiencing more influences of the magnetic field. The shell-crossing field is energy ($V_g$) dependent (dashed line in Fig. 2a or 3b) in the same tube as a result of non-linearity in *E(k)* with higher group velocity along the tube axis for states further away from the bandgap. Thus, the electrical transport data reveals detailed band structures of SWNTs including bandgap and non-linearity in *E(k)*.

Also interesting is that the high quality suspended *quasi-metallic* SWNTs exhibit a systematic trend in conductance level vs. SWNT bandgap. In an earlier suspended SWNT with $E_g$ ~ 14meV and high conductance of G ~ 2.7 $e^2/h$, we observed clear Fabry-Perot interference for p-channel transport.[9] For the two tubes in Dev1 and 2 with larger $E_g$ (60mV and 100 mV respectively), the p-channel conductance systematically decrease ($G$ ~ 1.5 $e^2/h$ and 1.1 $e^2/h$ respectively). These are attributed to higher and thicker Schottky barriers (SB) formed between Pt to the valence band of nanotubes with larger $E_g$ and hence higher resistance at the metal-tube junctions due to a thicker SB width for tunneling for transport at low temperatures (Fig. 4a). With increasing $E_g$ and lower $G$, the devices exhibit an evolution[19] from FB-type interference to shell-filling and then deep into Coulomb blockade (Fig. 4b). The electron addition energy $U_{eff}$ increases as $G$ level decrease, as seen here for the same nanotube in Dev1 in the p- and n-channels (Fig.1c) with $U_{eff}$=2.8meV and 3.9meV respectively (due to larger Schottky barrier height to the n- than the p-channel with Pt contacts).

We also investigated suspended semiconducting SWNTs with larger bandgaps ($E_g$ ~ 0.4eV) than the quasi-metallic ones. Devices of this type of SWNT were known to be p-type field effect transistors at $T$~300K with diminished $G$ under increasing $V_g$ (Fig. 4c for a $L$~500µm suspended tube). At low temperatures, the p-channel $G$ were typically very low (Fig. 4c, $G$ ~ 0.03 $e^2/h$, 2 orders of magnitude lower than at $T$~300K) due to quenched thermal activation over the large Schottky barrier formed at the Pt-SWNT contact junctions (for large $E_g$). Consistent with the trend in Fig. 4b, the p-channels of our suspended semiconducting SWNTs were deep in the CB regime at low temperatures with sharp Coulomb oscillations. Nevertheless, the CB features were highly regular (Fig. 4c) corresponding to a single coherent quantum dot without apparent disorder along the tube length $L$=560nm. Single-dot behavior has been observed in ~90% of our suspended SWNT ($L$<1μm) devices in the low conductance limit including ~30 semiconducting tubes.

Lastly, we present the electrical characteristics of suspended metallic SWNTs. True-metallic arm-chair SWNTs were rare with a low abundance of a few percent.[20] We did encounter one or two such nanotubes out of ~100 suspended devices studied. Fig. 4d shows such a $L \sim 2$ μm suspended metallic SWNT exhibiting little gate dependence (over a wide gate-voltage range) from room temperature down to $T$~120 K accompanied by metal-like increase in conductance at lower $T$ (Fig. 4d). Broad oscillation features were observed at $T$=2K over a high conductance background near $2e^2/h$. The lack of gate dependence over a wide temperature range indicates the high quality of the 2 μm long suspended tube, as defects and disorders are known to cause energy or gate-dependent resonance transport behavior observed for metallic tubes on substrates (at 300K).[21] We suggest that the as-grown suspended state of metallic SWNT is most likely to preserve the true metallic nature of arm-chair tubes. However, the irregular FP-like oscillations observed at 2K (Fig. 4d) does suggest imperfection in the 2 μm long suspended tube manifesting in its low-$T$ transport characteristics.

**Acknowledgements**

This work was supported by MARCO Focused Research Center on Materials, Structures and Devices and a NSF NIRT.

* Correspondence and request for materials should be addressed to HD, hdai@stanford.edu

**Competing financial interests**

The authors declare that they have no competing financial interests

**Figure Legends:**

**Figure 1**. Suspended SWNTs. (a) Schematic device with a local gate at the bottom of the trench. (b) Scanning electron microscopy (SEM) image of the actual device described in Fig. 3, scale bar is 0.5$\mu$m. (c) $G$-$V_g$ characteristics of a SWNT (Dev1) recorded at $T$=300 mK under $V$ =1mV and B=0T. Heights of the bars along the top-axis correspond to peak spacings $\Delta V_g$ (right vertical axis) between conductance peaks along the $V_g$ axis. A 4-peak shell is highlighted (by dashed lines) for the p-channel (negative $V_g$ side) and n-channel (positive $V_g$ side) respectively. (d) Energy dispersion $E(k)$ for the valence band. Quantization of wavevectors along the length of carbon nanotube ($k_n$=n$\pi$/L, n=integers) is indicated as evenly spaced vertical lines. Each $k_n$ gives rise to a shell (represented by the horizontal red levels) each comprised of 4 states corresponding to K and K' subbands and spin-up and spin-down. (e) Details of two of the shells in (c). 4-electrons fill each shell with a charging energy of $U_{eff}$. To reach the next shell, in addition to $U_{eff}$, energy difference between the quantized shells $\Delta_n$ needs to be paid.

**Figure 2**. A suspended quasi-metallic SWNT (same as Dev1 in Fig.1) in magnetic fields. (a) Experimental data of p-channel conductance $G$ (represented by color) vs. $V_g$ and magnetic field B (-8 T to 8T, parallel to tube axis) based on 160 $G$-$V_g$ curves from B = –8T to 8T in 0.1 T steps. The circles are drawn to highlight two 4-peak shells and the one at B ~ 5T is formed by shell-crossing. The dashed line marks the shell-crossing fields vs. energy or $V_g$. (b) Calculated energy $E$ levels (Eq. 2) for the electronic states vs. field B and energy to fit the experimental data in (a). The region framed by dashed green lines corresponds to Fig.2d. (c) Evolution of the K and K' subbands in a magnetic field and splitting of shells. (d) Evolution of the quantized K (red dashed lines) and K' (blue dashed lines) states in each shell as a function of B field according to Eq. 2 except for without including the Coulomb charging energy $U_{eff}$. Level crossing occurs as electrons fill in the lower energy states first. This diagram gives rise to the framed region in (b) after including 2-fold spin degeneracy and an effective charging energy $U_{eff}$ between each electron. Note: gate efficiency conversion factor (between $V_g$ and $E$) for this device is $\alpha \sim 0.015$.

**Figure 3**. A second suspended quasi-metallic SWNT (Dev2). (a) $G$-$V_g$ characteristics for the suspended SWNT recorded at $T$=300 mK under $V$ =1mV. Top: spacing between adjacent conductance peaks. (b) A plot of $G$ (represented by color) vs. $V_g$ and magnetic field B (-8 T to 8T) based on 160 $G$-$V_g$ curves from B = –8T to 8T in 0.1 T steps. The dashed line marks the shell-crossing fields vs. $V_g$. The n-channel of this device (not shown) exhibits low conductance level and Coulomb blockade. (c) Calculated energy $E$ levels for the electronic states vs. B and energy to fit the experimental data in (b). Certain discrepancy with experimental data in (b) is seen near the band edge since the shell-filling model does not apply to states very close to the band edge. Note: gate efficiency conversion factor (between $V_g$ and $E$) for this device is $\alpha \sim 0.032$.

**Figure 4**. Contacts and properties of various types of suspended nanotubes. (a) Band diagrams showing a higher Schottky barrier (responsible for higher contact-junction resistance) formed to the p-channel (valence band) of a larger bang-gap SWNT (lower panel) than that of a smaller band-gap tube (top-panel) at the metal-tube contact. (b) A schematic drawing to illustrate the crossover from Fabry Perot (FP) interference, to shell-filling (2$^{nd}$ and 3$^{rd}$ curves, the 2$^{nd}$ one has higher conductance and is over a broad FP-like peak) and Coulomb blockade (CB) with increasing charging energy ($U_{eff}$) as the conductance level of a nanotube device decreases due to higher contact-junction resistance for larger band-gap nanotubes contacted by Pt. (c) $G$-$V_g$ characteristics of a semiconducting suspended SWNT at 2K and room temperature (inset). (d) $G$-$V_g$ characteristics of a true-metallic suspended SWNT at various temperatures indicated.

**Figure 1**

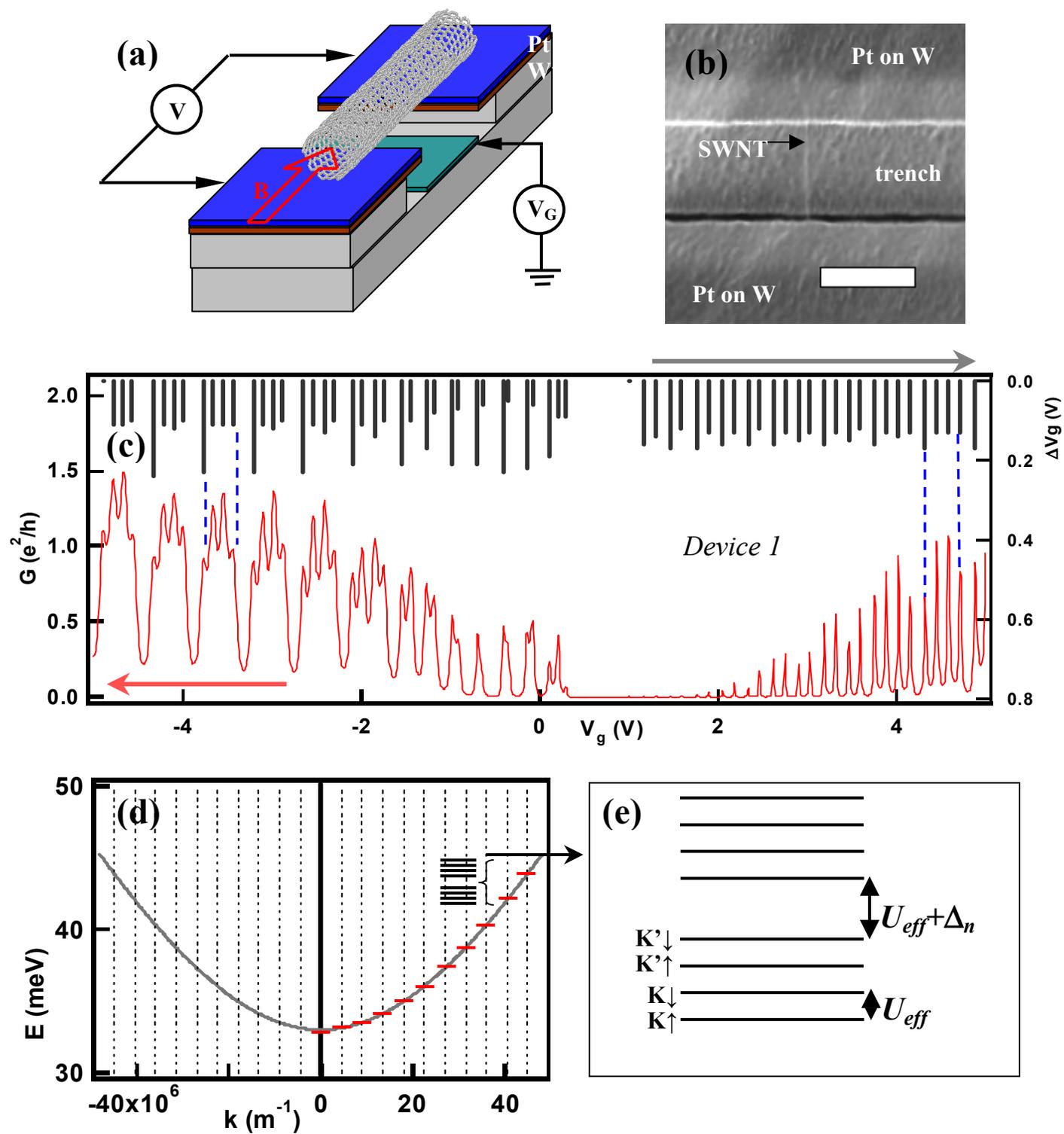

**Figure 2**

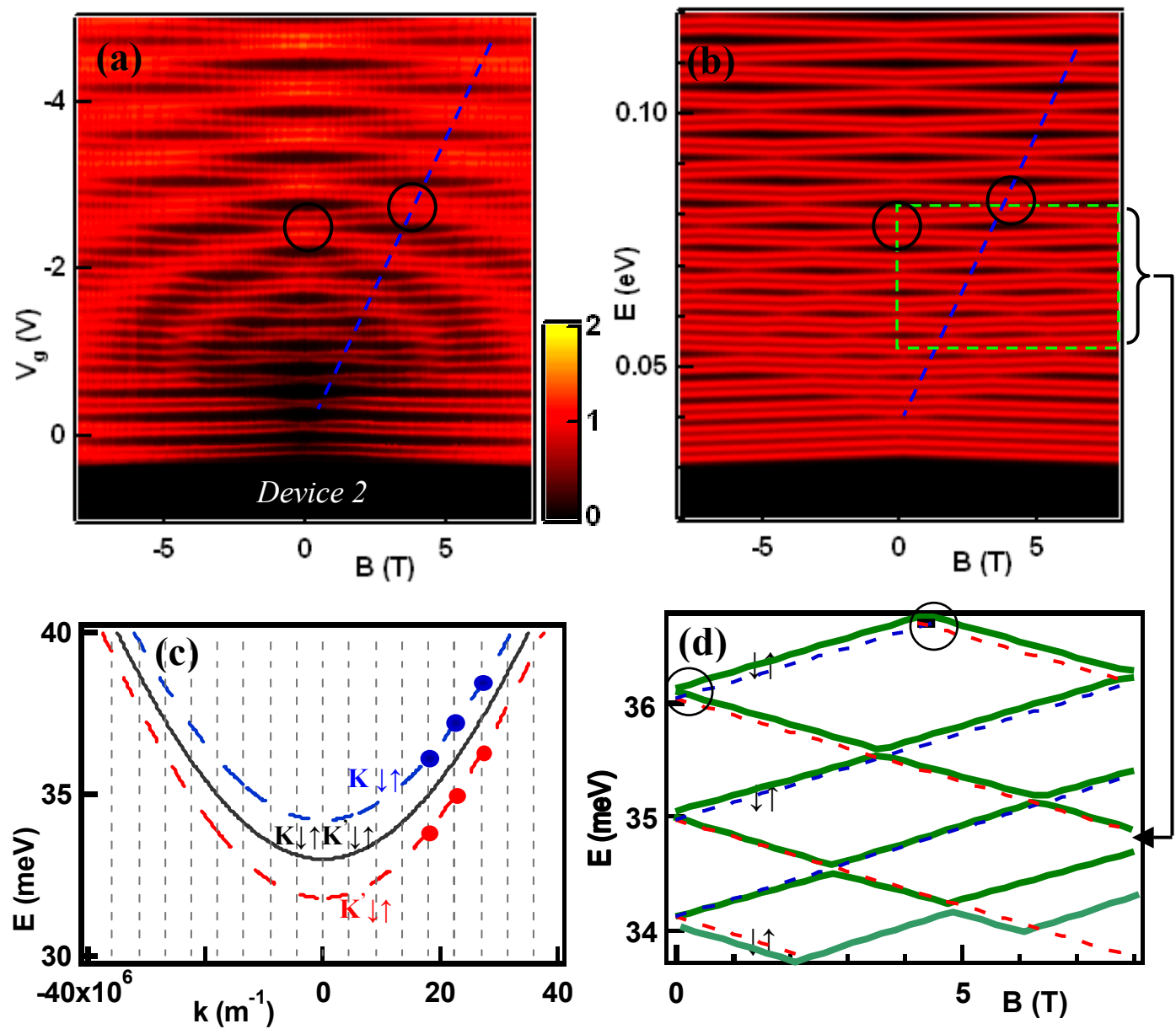

**Figure 3**

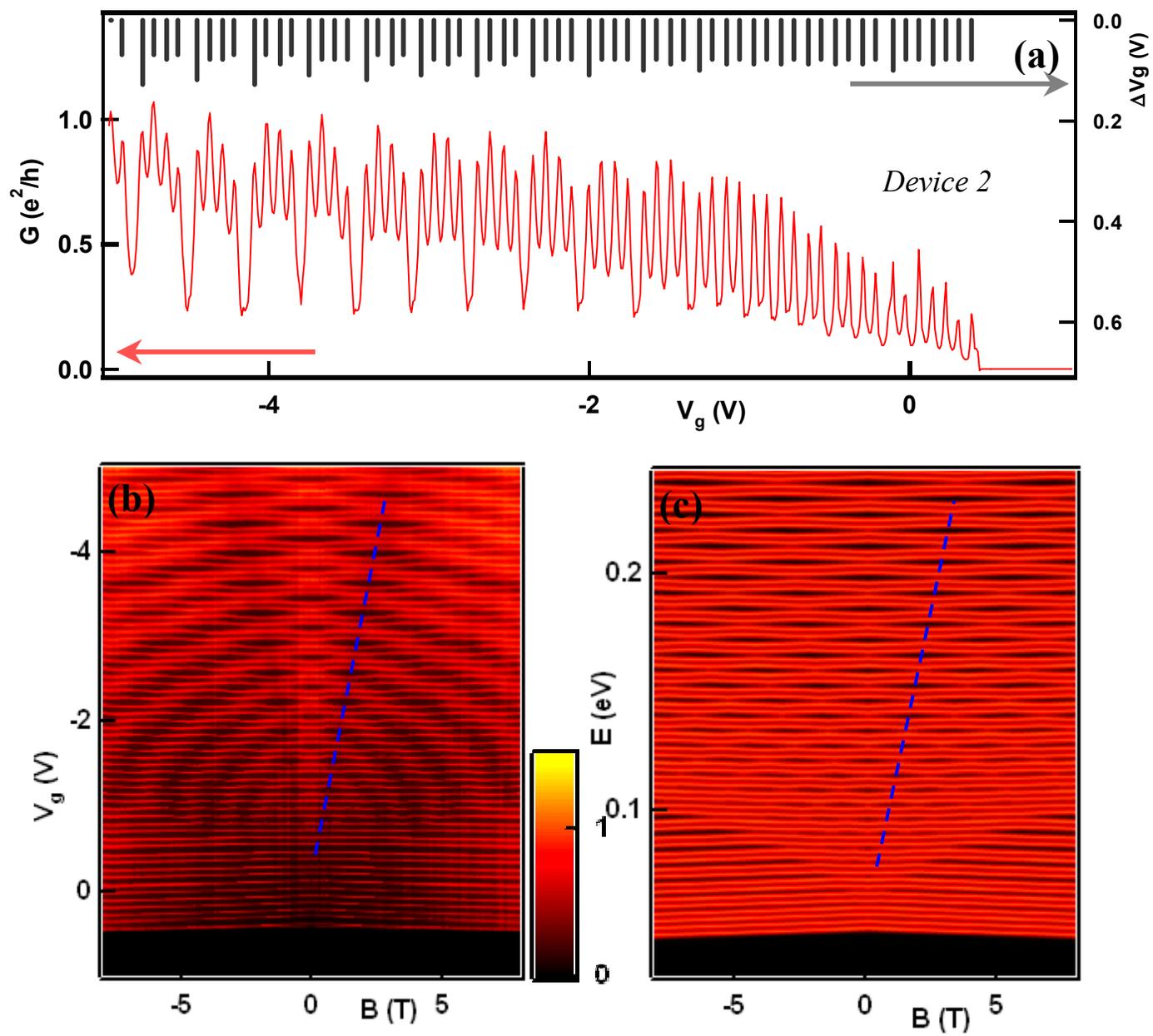

**Figure 4**

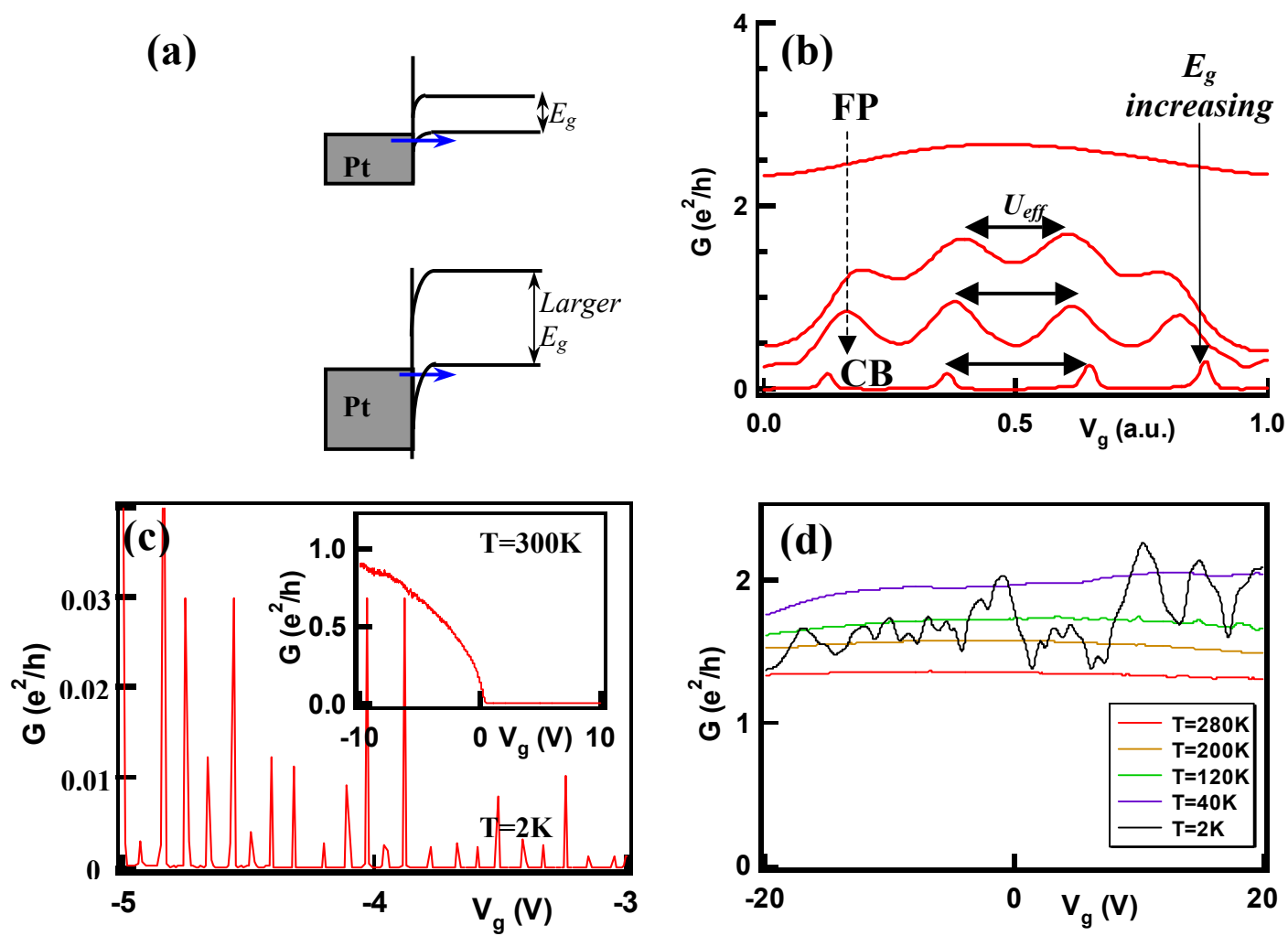